\documentclass{aastex632}

\usepackage{hyperref}
\usepackage[version=3]{mhchem}
\usepackage{threeparttable}
\usepackage{color}

\newcommand*\mearth{\mathrm{M}_\oplus}
\newcommand*\rearth{\mathrm{R}_\oplus}

\newcommand{\sub}[1]{_{\mathrm{#1}}}
\newcommand{\appropto}{\mathrel{\vcenter{
  \offinterlineskip\halign{\hfil$##$\cr
    \propto\cr\noalign{\kern2pt}\sim\cr\noalign{\kern-2pt}}}}}

\definecolor{nncol}{rgb}{0.0,0.3,0.9}

\accepted{May 13, 2025}

\begin{document}

\title{Evolution of steam worlds: energetic aspects}

\author[0000-0002-8949-5956]{Artyom Aguichine}
\affiliation{Department of Astronomy and Astrophysics, University of California, Santa Cruz, CA, USA}
\email{aaguichi@ucsc.edu}

\author[0000-0002-7030-9519]{Natalie Batalha}
\affiliation{Department of Astronomy and Astrophysics, University of California, Santa Cruz, CA, USA}

\author[0000-0002-9843-4354]{Jonathan J. Fortney}
\affiliation{Department of Astronomy and Astrophysics, University of California, Santa Cruz, CA, USA}

\author[0000-0002-1608-7185]{Nadine Nettelmann}
\affiliation{Department of Astronomy and Astrophysics, University of California, Santa Cruz, CA, USA}

\author[0000-0002-4856-7837]{James E. Owen}
\affiliation{Imperial Astrophysics, Department of Physics, Imperial College London, Prince Consort Road, London SW7 2AZ, UK}
\affiliation{Department of Earth, Planetary, and Space Sciences, University of California, Los Angeles, CA 90095, USA}

\author[0000-0002-1337-9051]{Eliza M.-R. Kempton}
\affiliation{Department of Astronomy, University of Maryland, College Park, MD 20742, USA}



\begin{abstract}

Sub-Neptunes occupy an intriguing region of planetary mass-radius space, where theoretical models of interior structure predict that they could be water-rich, where water is in steam and supercritical state. Such planets are expected to evolve according to the same principles as canonical \ce{H2}-He rich planets, but models that assume a water-dominated atmosphere consistent with the interior have not been developed yet. Here, we present a state of the art structure and evolution model for water-rich sub-Neptunes. Our set-up combines an existing atmosphere model that controls the heat loss from the planet, and an interior model that acts as the reservoir of energy. We compute evolutionary tracks of planetary radius over time. We find that planets with pure water envelopes have smaller radii than predicted by previous models, and the change in radius is much slower (within $\sim$10\%). We also find that water in the deep interior is colder than previously suggested, and can transition from plasma state to superionic ice, which can have additional implications for their evolution. We provide a grid of evolutionary tracks that can be used to infer the bulk water content of sub-Neptunes. We compare the bulk water content inferred by this model and other models available in the literature, and find statistically significant differences between models when the uncertainty on measured mass and radius are both smaller than 10\%. This study shows the importance of pursuing efforts in the modeling of volatile-rich planets, and how to connect them to observations.

\end{abstract}

\keywords{Planet interiors -- Evolution -- Interior structure modeling -- Water worlds}


\section{Introduction} \label{sec:intro}

The search for water rich exoplanets represents an overarching goal for astrobiology and the search for life. In the Solar System, there is a line of evidence suggesting that icy moons of giant planets (Europa, Ganymede, Titan, Enceladus, etc.) have extensive water oceans in their interior \citep{Lunine2017,Vance2018,Journaux2020}. Pluto, Uranus, and Neptune are also believed to be water rich, supposedly formed from ice-rich building blocks beyond the water iceline \citep{Mousis2018_pss,Malamud2024}. The most striking evidence for the existence of deep sub-surface oceans are the water-dominated plumes expelled from the internal ocean of Saturn's moon Enceladus \citep{Waite2009,Hansen2011}. These observations motivated the development of interior structure models to determine the bulk water content using masses, radii, the Love number $k_2$, and the moment of inertia for icy moons \citep{Sotin2007,Jin2012,Journaux2020,Trinh2023}.

Over the past decade, extensive efforts were dedicated to the characterization of exoplanets. Precise measurements of exoplanet masses and radii can be used to constrain their bulk compositions using interior models \citep{Seager2007,Fortney2007,Valencia2007,Zeng2013,Zeng2016,Zeng2019}. Sub-Neptunes, exoplanets with radii between $\sim$1.8 to $\sim$3.5 $\rearth$, are of particular interest because their densities are compatible with planets that have water rich interiors \citep{Zeng2019,Mousis2020_iop,Luque2022}, although their radii can also be explained by \ce{H2}-He envelopes that account for at most a few percents of the planet mass \citep{Fortney2007,Lopez2014,Rogers2023}.

NASA's \textit{Kepler} space telescope discovered more than 1000 such planets\footnote{NASA Exoplanet Archive, July 2024 \url{https://exoplanetarchive.ipac.caltech.edu/}}, allowing the characterization of the interiors of sub-Neptune, the most common type of exoplanet discovered to date, on a population level. Another major finding of the \textit{Kepler} mission is a gap in the distribution of planetary radii that separates super-Earths ($\leq 1.8 \rearth$) from sub-Neptunes ($\geq 1.8 \rearth$) \citep{Fulton2017,Parc2024,Schulze2024}. This feature had been predicted \citep{Lopez2013,Owen2013} and is reproduced \citep{Rogers2023} by theoretical models of atmospheric mass-loss, assuming that the interiors of sub-Neptunes are made of a rocky core with a \ce{H2}-He envelope on-top. 
More recent studies aim to reproduce the radius gap using planet synthesis models, which incorporate prescriptions for initial volatile budget from formation models, atmospheric evolution, and interior structure to track radius evolution. These studies find that the radius gap can be explained by a dichotomy in bulk composition, with super-Earths mostly being rocky and sub-Neptunes water-rich \citep{Izidoro2022,Burn2024,Chakrabarty2024,Venturini2024}. However, these conclusions heavily rely on interior structure models that predict theoretical planetary radii from a given composition, which are then compared to measurements. For this reason, understanding the physics at play in the interiors of water worlds is critical for relating composition to bulk planet properties in sub-Neptunes.

Due to the proximity to their host star, known sub-Neptunes tend to have high equilibrium temperatures. In fact, most of them\footnote{Almost 97\% using the criteria from \cite{Turbet2019}.} are beyond the runaway greenhouse limit \citep{Nakajima1992,Kopparapu2013,Turbet2019}, meaning that they cannot maintain liquid water oceans. Instead, such planets would have steam atmospheres on top of supercritical water envelopes \citep{Mousis2020}. These considerations led to the development of coupled atmosphere-interior models to compute reliable mass-radius relationships for such planets \citep{Turbet2020,Aguichine2021}. Although steam atmospheres have masses that are negligible compared to the bulk water content, they account for a significant portion of the total planetary radius \citep{Turbet2020,Vivien2022}. In addition to this, water in a supercritical state has a much lower density than it does in the liquid state. Consequently, envelopes of supercritical water are even more extended than planets with deep liquid oceans (and eventual high-pressure ices). Updated mass-radius relations changed the inferred bulk compositions of sub-Neptunes, since the radii of sub-Neptunes are now compatible with interiors where envelopes are dominated by water in a supercritical state. This picture aligns with recent estimates of atmospheric compositions with JWST, where atmospheres of most sub-Neptunes so far exhibit metallicities greater than at least 100 times the solar value \citep{Kempton2023-gj1214b,Alderson2024,Beatty2024,Wallack2024,Benneke2024}. For K2-18b, \cite{Madhusudhan2023} report that the volume mixing ratios of \ce{CH4} and \ce{CO2} are both $\sim 1\%$, which corresponds to a C/H ratio of $\sim 50 \times$ the solar value.

Contrary to refractory materials, the volume density of volatile compounds is highly sensitive to temperature. Therefore, the temperature profile in a planetary interior has a significant impact on the planet's radius when the interior is dominated by volatiles. The dependence of the planetary radius on temperature is sufficiently high to impact the inferred bulk composition, as shown in the case of Jupiter \citep{Helled2022} and Uranus \citep{Neuenschwander2024}. One of the processes that determines the temperature in planetary interiors is the gradual release of internal heat, acquired from formation \citep{Hubbard1977}. The cooling of the interior is associated with planet-wide contraction and thus a decrease in planetary radius over time. This effect has been extensively studied in the context of metal-rich, H-He dominated envelopes \citep{Fortney2003,Fortney2007,Mordasini2012b,Lopez2014,Thorngren2016}, but most interior models with water dominated atmospheres are static, i.e. the thermodynamic state of water is assumed \citep{Zeng2019,Madhusudhan2020,Huang2022}. Only a few studies modeled the interior structure of planets with pure steam atmospheres and with \citep{Nettelmann2011} or without \citep{Haldemann2024} thermal contraction; however, these models are not consistent in regard of composition of the atmosphere and the interior model. For instance, \cite{Nettelmann2011} incorporate age in the same way as this work does but assumes a $50\times$ Solar metallicity atmosphere, and \cite{Haldemann2024} uses the age-luminosity relation that was calibrated for a $1\times$ Solar metallicity atmosphere \citep{Mordasini2020}.

The purpose of this study is to determine what role does the axis of time plays in controlling the radius of water-rich sub-Neptunes, which ultimately determines our interpretation of their bulk composition. We investigate energetic aspects of steam worlds by modeling their thermal history using a compositionally consistent coupled atmosphere-interior model. We couple the interior model of steam worlds from \cite{Aguichine2021} with the recently published atmosphere model adapted to pure steam atmospheres by \cite{Kempton2023}. Section \ref{sec:methods} describes the model and the numerical procedure to model the evolution. In Section \ref{sec:results} we show modeled atmosphere-interior structures, and make a comparison with the work of \cite{Nettelmann2011} for the case of GJ 1214 b. We summarize our findings in Section \ref{sec:ccls}. Understanding how steam worlds evolve is key to understand sub-Neptunes on a demographic level, possibly breaking the degeneracy in their bulk compositions using the system age measurement by Gaia and PLATO.

\section{Methods} \label{sec:methods}

In this section, we describe how the new grid of atmosphere models of \cite{Kempton2023} was implemented into the static model of \cite{Aguichine2021}, and how the dimension of time was added. We review the main features of both models. The procedure developed here to model the evolution of water worlds is heavily inspired by the work of \cite{Fortney2007,Nettelmann2011, Lopez2012} and \cite{Lopez2014}.

\subsection{Interior model} \label{sec:interior-model}

The interior structure model, based on the work of \cite{Brugger2016,Brugger2017}, is thoroughly described in \cite{Aguichine2021}. For the sake of keeping our new results comparable with the older static model, we do not make any changes to the interior part of the model, and only review details that are relevant to the time-evolution modeling.

Our interior model iteratively solves the equations describing the interior of a planet in 1D:

\begin{eqnarray}
\frac{\mathrm{d} g}{\mathrm{d} r}&=&4 \pi G \rho-\frac{2 G m}{r^{3}}, \label{eq:gauss}\\
\frac{\mathrm{d} P}{\mathrm{d} r}&=&-\rho g, \label{eq:hydrostatic}\\
\frac{\mathrm{d} T}{\mathrm{d} r}&=&-g\gamma T\frac{\mathrm{d} \rho}{\mathrm{d} P}, \label{eq:temp_grad}\\
P &=& f(\rho,T), \label{eq:solve_eos}
\end{eqnarray}

\noindent where $g$, $P$, $T$ and $\rho$ are gravity, pressure, temperature and density profiles, respectively. $m$ is the mass encapsulated within the radius $r$, $G$ is the gravitational constant, and $\gamma$ is the Gr\"uneisen parameter. Five distinct layers are considered in our interior model:

\begin{itemize}
	\item a core made of metallic Fe and FeS alloy;
	\item a lower mantle made of bridgmanite and periclase;
	\item an upper mantle made of olivine and enstatite;
	\item an ice VII phase;
	\item an envelope covering the whole fluid region of \ce{H2O}.
\end{itemize}

The bulk composition of the interior is determined by the water mass fraction (WMF) noted $x_{\mathrm{H_2O}}$, and the iron core mass fraction (CMF) noted $x_{\mathrm{core}}$. For any given value of $x_{\mathrm{H_2O}}$, we adapt the value of $x_{\mathrm{core}}$ so that the composition of the refractory part (not including the water envelope) is that of the Earth \citep[32.5\% iron core, and 67.5\% of silicate mantle][]{Sotin2007}. This can be justified by the fact that if the planet accreted a huge reservoir of water from beyond the iceline, the refractory content of the planet should not differ from the solar composition \citep{Drazkowska2017,Aguichine2020,Miyazaki2021}. This hypothesis is supported by the fact that the composition of rocky planets and comets show no strong deviations from the solar composition \citep{McCoy2005,Stacey2005,Sotin2007,Lodders2009,Luque2022}. For all layers except the envelope, we use the equation of state (EOS) of Vinet with a Debye thermal correction  \citep{Vinet1989}, with parameters from \cite{Stacey2005} and \cite{Sotin2007} \citep[see][for a summary]{Aguichine2021}. For the \ce{H2O} envelope, we use the EOS of \cite{Mazevet2019}. At every P-T point in the interior, the code determines whether \ce{H2O} is in the ice VII or fluid phase
according to the phase transition curve of \cite{Wagner2011}. For the fluid phase we use the 
\cite{Mazevet2019} water EOS is while for the ice Vii phase the Vinet EOS. The temperature profile in all layers is adiabatic, and we denote $s$ the specific entropy of the envelope (here, made of pure \ce{H2O}), which is constant throughout the envelope by definition\footnote{The code provided in the published version of \cite{Mazevet2019} contains a typo in the formula used for the computation of the entropy (but not its derivatives). Thus when the derivatives are used to compute the adiabat, the EOS produces correct adiabatic P-T profiles, but values of entropy yield a non-constant specific entropy profile. A corrected version can be downloaded from the author's website: \url{http://www.ioffe.ru/astro/H2O/index.html}. As of August 2024, the AQUA EOS \citep{Haldemann2020} tabulated the entropy with the non-corrected version of the \cite{Mazevet2019} EOS. Therefore, we do not use it here.}. We neglect temperature shifts between different layers that are traditionally used to mimic diffusive heat transfer \citep{Valencia2006,Sotin2007} since the change in radius for sub-Neptunes is dominated by the envelope made of volatiles, rather than the thermal state of the refractory core \citep{Stacey2005,Valencia2006,Vazan2018}.

Apart from the compositional parameters $x_{\mathrm{H_2O}}$ and $x_\mathrm{core}$, the interior model takes as input the planet mass $M_\mathrm{b}$, surface pressure $P_\mathrm{b}$ and surface temperature $T_\mathrm{b}$ to compute the interior structure. In turn, this gives the planet radius $R_\mathrm{b}$ at $P_\mathrm{b}$, from which the surface gravity $g_\mathrm{b}$ can be obtained. The subscript $b$ refers to quantities at the boundary between the interior model and the atmosphere model, which is chosen at $P_\mathrm{b}=1000$ bar.

\subsection{Atmosphere model grid} \label{sec:atmosphere-model}

\cite{Kempton2023} present a large grid of atmospheric structures for a wide variety of parameters: \ce{H2O} volume mixing ratio, surface gravity, intrinsic temperature, equilibrium temperature, and host star type; and associated transmission spectra. In this work, we limit ourselves to pure \ce{H2O} atmospheres, although the method presented here can be generalized to interiors where the atmosphere and envelope are mixtures of \ce{H2O} with \ce{H2}-He.

The atmospheric structures are generated using the open-source package \texttt{HELIOS} \citep{Malik2017,Malik2019a,Malik2019b}. The one-dimensional P-T profile of each atmosphere is computed in radiative-convective equilibrium, assuming that heat is perfectly redistributed over both hemispheres (heat redistribution factor of 0.25).

Convection in the atmosphere is treated via convective adjustment. This requires knowledge of the adiabatic coefficient and the specific heat coefficient of the atmosphere's constituents. For \ce{H2O}, these are computed using the IAPWS-95 formulation \citep{Wagner2002}, which includes non-ideal behavior of \ce{H2O}.

The opacity of \ce{H2O} is computed using the open-source code \texttt{HELIOS-K} \citep{Grimm2015,Grimm2021} using the POKAZATEL line list \citep{Polyansky2018}. The spectra are integrated with a high resolution of 0.01 cm$^{-1}$, assuming a Voigt profile for each spectral line shape. Lastly, the opacity of water also includes Rayleigh scattering \citep{Cox2000,Wagner2008}. When computing the Rosseland mean opacity of \ce{H2O} \cite{Grimm2021} used a fixed line-wing cutoff at 100 cm$^{-1}$, which leads to \ce{H2O} opacity decreasing as pressure increases, which is unphysical. \cite{Kempton2023} re-computed the Rosseland mean opacity by applying a wing cutoff at 500 cm$^{-1}$ when pressures are greater than 100 bar, and recovered the expected trend of increasing Rosseland mean opacity with pressure (see their Appendix A).

To model planets orbiting a solar-type star, the stellar flux is approximated by a blackbody at a temperature $T_{\star} = 5800$ K and a solar radius. For M-type stars, the stellar spectra is generated with \texttt{PHOENIX} \citep{Husser2013} at $T_{\star} = 3026$ K and other parameters representative of GJ 1214 \citep{Harpsoe2013}. In \texttt{HELIOS}, the specified equilibrium temperature (here, 500 K or 700 K) is achieved by setting the orbital distance that would result in the correct value of $T_{\mathrm{eq}}$, assuming zero albedo and full planet heat redistribution.

The grid of atmospheres by \cite{Kempton2023} is provided in the form of pressure, temperature, altitude profiles. The parameter space of the grid contains 15 points for surface gravities $\log g\sub{b} = [0.0, 1.7]$ (SI), 14 points for intrinsic temperature $T_{\mathrm{int}} = [10,400]$ K, 2 points for equilibrium temperatures $T_{\mathrm{eq}} = [500,700]$ (K), and a Sun-like or M-type host star, making a total of 840 atmospheric profiles. All atmospheric profiles provide temperature and altitude on the same grid of pressures. Intermediate profiles are thus interpolated using 3D-linear interpolation, and outside of the grid profiles are linearly extrapolated. We only interpolate in dimensions of $\log g\sub{b}$, $T_{\mathrm{int}}$, and $T_{\mathrm{eq}}$, and choose whether the host is Sun-like or M-type. From each profile, we extract the 
temperature at the bottom of the atmosphere $T_{\mathrm{b}}$, which is fixed at a pressure of $P_{\mathrm{b}}=1000$ bar. This provides us a secondary grid of surface temperature $T_{\mathrm{b}}$ computed as a function of $\log g\sub{b}$ and $T_{\mathrm{int}}$, 
which is similar to the method employed in \cite{Fortney2007,Nettelmann2011,Lopez2014}.
From the altitude profiles, we also extract the thickness of the atmosphere $R_{\mathrm{atm}}$ as the altitude at the transiting pressure $P_{\mathrm{tr}}$. Here, we choose $P_{\mathrm{tr}}=1$ $\mu$bar (0.1 Pa) for the sake of comparison with the \cite{Aguichine2021} model, although the transiting radius for a cloud-free atmosphere could also be located much deeper, at levels of $1-100$ mbar ($10^2-10^4$ Pa). The transiting radius can be computed by integrating the optical depth along an incident chord, but this calculation will depend on the spectral band in which the planet is observed. For example, the radius of a planet can change significantly when measured with Kepler (0.4--0.9 $\mu$m) or HST's WFC3 G141 filter (1.1--1.7 $\mu$m), as shown by \cite{Gao2020}. For this reason, we compute the physical radius at 20 mbar and 1 $\mu$bar. The radius at 20 mbar is roughly the transiting pressure of a clear steam-dominated atmosphere at 1 $\mu$m \citep{Nettelmann2011,Grimm2018,Piaulet2023,Tang2024}. The radius at 1 $\mu$bar corresponds to the pressure level where photochemical hazes are expected to form for the typical equilibrium temperature in sub-Neptunes \citep{Kawashima2019,Lavvas2019,Gao2020-hazes}. We thus recommend using the radius at $P_{\mathrm{tr}}=1$ $\mu$bar, which is the value that will be used throughout this paper. In our model, the radius can be read at any pressure level using the profiles from \cite{Kempton2023}, but when a grid of radii are provided at 20 mbar and 1 $\mu$bar the radius at any other pressure level can also be derived by assuming the atmosphere is isothermal and hydrostatic. In all of our parameter space, the ratio between the atmospheric thickness at 1 $\mu$bar and 20 mbar is between 1.27 and 1.63 for all atmospheres considered in this study. Most sub-Neptunes have surface gravities such that $\log g\sub{b} \ge 0.7$, a parameter space where the atmosphere thickness at $1$ $\mu$bar is at most 0.44 $\rearth$. This means that there is at most $\sim 0.12 ~\rearth$
of difference in atmosphere thickness between the two methods, but in most cases this difference is of order of $\sim 0.06 ~\rearth$ (where $\log g\sub{b} = 1.0$). We note that the altitude profile was computed by integrating the hydrostatic equilibrium assuming constant surface gravity. With this assumption, the atmosphere thickness may be underestimated, especially for planets at lower masses. We make an attempt at quantifying this difference using Eq. \ref{eq:epsilon-gravity}. The total planetary radius is obtained by summing the radius computed in the interior model 
$R\sub{b}$ with the atmosphere thickness $R_{\mathrm{atm}}$ between 1000 bar and $1$ $\mu$bar.

\subsection{Thermal evolution} \label{sec:coupling-method}

The interior and atmosphere models are connected at a surface pressure $P_{\mathrm{b}}=1000$ bar. Interior and atmosphere models can be connected at any arbitrary pressure as long as the physics describing the bottom of the atmosphere is identical to the top of the interior model. For coupled evolution models such as the one presented here, the requirement is that the bottom of the atmosphere is fully convective (i.e. adiabatic). For this reason, atmospheric profiles need to be generated down to surface pressures that are as high as possible to satisfy this condition. Surprisingly, even at 1000 bar, the atmospheres are rarely adiabatic at the bottom. We will comment the consequences of this inconsistency in the next section.

The time evolution is computed by integrating numerically the energy balance equation \citep{Lopez2012,Lopez2014}:
\begin{equation}
    L_{\mathrm{int}} = - \frac{\mathrm{d} E_{\mathrm{int}}}{\mathrm{d} t}, \label{eq:energy-balance}
\end{equation}
where $L_{\mathrm{int}}=4\pi R_{\mathrm{b}}\sigma_{\mathrm{SB}} T_{\mathrm{int}}^4 $ is the heat loss from he interior, and $dE_{\mathrm{int}}$ is the change of total energy in the interior with $\sigma_{\mathrm{SB}}$ the Stefan-Boltzmann constant 
At any given time, the rate of change in internal energy is given by:
\begin{equation}
    \frac{\mathrm{d} E_{\mathrm{int}}}{\mathrm{d} t} = \int_{\mathrm{env}} T\frac{\partial s}{\partial t} \mathrm{d} m + \int_{\mathrm{core}} c_{\mathrm{v}} \frac{\mathrm{d} T_{\mathrm{core}}}{\mathrm{d} t} \mathrm{d} m + \frac{\mathrm{d} E_{\mathrm{grav, core}}}{\mathrm{d} t} + L_{\mathrm{radio}}, \label{eq:energy-rate}
\end{equation}
where the 4 terms on the right hand side account for cooling of the envelope, heat extracted from the core (iron core and mantle), gravitational energy, and energy produced by radiogenic heating in the interior, respectively. In the envelope, the energy is computed by integrating $Ts\mathrm{d}m$, which automatically accounts for both the thermal energy due to cooling and the gravitational energy due to contraction. For the core, we use a simpler approach and only consider the energy associated with the temperature decrease.
Following \cite{Gonzales-Cataldo2023}, we use $c_{\mathrm{v}} = 5.11~k_{\mathrm{B}}/\mathrm{atom}$, where $k_{\mathrm{B}}$ is the Boltzmann constant, which corresponds to the specific heat of liquid iron around 10 Mbar (1 TPa). \cite{Gonzales-Cataldo2023} also find $c_{\mathrm{v}} = 4.94~k_{\mathrm{B}}/\mathrm{atom}$ and $c_{\mathrm{v}} = 5.64~k_{\mathrm{B}}/\mathrm{atom}$ for solid and liquid iron, respectively, at pressures between 40 and 50 Mbar (4 and 5 TPa). These values agree with the experiments of \cite{Kraus2022} who found $c_{\mathrm{v}} = 4.2\pm1.0~k_{\mathrm{B}}/\mathrm{atom}$ at similar conditions, and also the values computed from Eq. (6) in \cite{Ichikawa2014}. We use the same specific heat for the iron core and the silicate mantle. Although the core can make up a significant portion of the total energy in the interior of a planet, most of the energy flux comes from the envelope, even at WMF$\sim 10\%$, or from radiogenic heating. The gravitational energy released from the core is accounted for manually:
\begin{equation}
    E_{\mathrm{grav, core}} = -G \int_{\mathrm{core}}  4 \pi r \rho(r) m(r) dr.
\end{equation}
The last energy source in the interior is the heat produced by the radioactive decay of heavy atomic nuclei, for which we use $L_{\mathrm{radio}}(t) = L_0*\exp(-t/\tau)$, where $L_0 = 270$ TW$\times (M\sub{p}/1~\mearth)$ and $\tau = 1.85$ Gyr \citep{Nettelmann2011}. This function is compatible with the present-day total radiogenic heat produced by Earth of 22 TW \citep{ONeill2020,Nimmo2020}.


The numerical procedure for our integration is similar to the one presented in \cite{Fortney2004PhD}. The planet is assumed to experience a "hot start" where the entropy of the envelope is initialized to a value of $s_0 = 10~k_{\mathrm{B}}$/baryon ($16.52$ kJ.K$^{-1}$.kg$^{-1}$). Based on this initial value and our choice $P\sub{b}=1000$ bar, we obtain the corresponding temperature $T\sub{b}$. $P\sub{b}$ and $T\sub{b}$ are used as boundary conditions (conditions \textit{"at the surface"}) to compute the planet's interior structure as described in Section \ref{sec:interior-model}. This interior structure gives the planetary radius, and the total energy in the interior $E\sub{int}$ (envelope, mantle, and core) for the initial entropy $s_0$. Additional details on the numerical procedure to integrate Eq. \ref{eq:energy-balance} are given in Appendix \ref{sec:appendix-evolution-scheme}. In our simulation, $t$ represents the age of the star as reported in the literature, meaning that the evolution must start at a earlier time $t_0$. In principle, $t_0$ represents the time at which the planet forming disk dissipates and the planet reaches its final orbit \citep{Strom1989,Hernandez2007}, when the planet is cooling with a constant incident flux. The initial time is believed to be somewhere between 1 and 10 Myr, which is the typical time for disk dissipation, but its actual value is unknown and it may differ from one planet to another. In reality, for 1D cooling models such as this one, $t_0$ must be chosen using physical arguments. The choice of $t_0$ is dependent on the adopted value of initial entropy $s_0$, since starting with a higher value of $s_0$ is equivalent to model the evolution starting from a smaller $t_0$. This is known as the hot start vs cold start choice, and it affects the early evolution of planets, although the evolution tracks eventually merge \citep{Marley2007,Spiegel2014}. Our choice for $t_0$ will be justified in the next section.




After the initialization, the atmosphere-interior structure is computed for an entropy $s-\Delta s$, and equations (\ref{eq:energy-balance}-\ref{eq:energy-rate}) are used to solve for the time-step $\Delta t$ that separates both states. If conditions on numerical precision are not satisfied, the computed atmosphere-interior structure is rejected, and the entropy step $\Delta s$ is adjusted accordingly.

The time evolution stops either i) when the age of the planet reaches 20 Gyr or ii) if the planet reaches an intrinsic temperature $T\sub{int}=10$ K. Although present-day Jupiter heat flux excess is $T\sub{int}=107.2$ K \citep{Li2018}, small values such as $T\sub{int}=10$ K are a natural outcome for planet of smaller masses. The second case corresponds to a situation where we reach the limit of the grid of atmospheres, and cannot compute the evolution further. Although this situation can occur relatively early ($\sim 10$ Gyr), when a planet reaches $T\sub{int}=10$ K the contraction is slow enough to become negligible. Therefore, we artificially continue the evolution curve to 20 Gyr by imposing the same radius as the previous time step. Once the simulation reaches 20 Gyr, we perform one last computation: an estimate of the time to full exhaustion. The planet is considered energetically exhausted when it has reached thermal equilibrium with the atmosphere, and no more heat is evacuated from its interior ($T\sub{int}=0$ K). Full exhaustion is characterized by an interior that is isothermal down to the center of the planet, and the temperature is that of the secondary isothermal region in the atmosphere. In reality, planets can only tend exponentially to this state without ever reaching it, since $T\sub{int}$ gradually decreases as the interior loses its heat. In our simulation, we estimate the exhaustion timescale by i) computing the energy of an isothermal interior, ii) determining the energy difference from the current state that needs to be evacuated to the exhausted state $\Delta E\sub{0K}$ and iii) dividing this energy difference by the intrinsic luminosity at 20 Gyr:
\begin{equation}
    \Delta t\sub{0K} = \frac{\Delta E\sub{0K}}{L\sub{int} (20~\mathrm{ Gyr})}. \label{ref:time-exhaust}
\end{equation}

\section{Results} \label{sec:results}

\subsection{Comparison with previous work} \label{sec:results-compare}

In this section, we compare our model, hereafter A25, with a pure water envelope model (case IIb) from \cite{Nettelmann2011}, denoted N11, and a previous model from \cite{Aguichine2021}, denoted A21 for the sub-Neptune GJ1214b. We adopt the same planet properties as in \cite{Nettelmann2011}, which are $M\sub{p}=6.55 ~\mearth$, $R\sub{p}=2.678 ~\rearth$, $T\sub{eq}=555$ K \citep{Charbonneau2009}, $x_{\mathrm{H_2O}}=  0.97$, $x_{\mathrm{core}^\prime}=0.325$, an M-type host star, which has importance for radiative transfer in the atmosphere, and a stellar age of $\tau=$3--9 Gyr.  

While  A25 and N11 are similar in methodology, they differ in the EOSs used and the atmosphere opacity. The A25 model is initialized with an initial specific entropy  
$s_0=10~k_{\mathrm{B}}$/baryon, and the planet radius is integrated forward. Also in the N11 model, the initial entropy is chosen high enough so that so that the planet cools quickly when young and the long-term evolution is unaffected by the initial value. Over time, the entropy must decrease, which is achieved by moving the RCB inward to higher pressures $P_{ad}$. The evolution is terminated when the planet radius has reached the prescribed value $R_p$, yielding a certain value $P_{ad}$ for the present state $t=\tau$. Different assumed WMFs can be realized by different values of $P_{ad}$ for the present state. For an imposed cooling time $\tau$=9 Gyr, the N11 models has $P\sub{ad}=300$ bar. 

\begin{figure}[!h]
    \centering
    \includegraphics[width=0.5\linewidth]{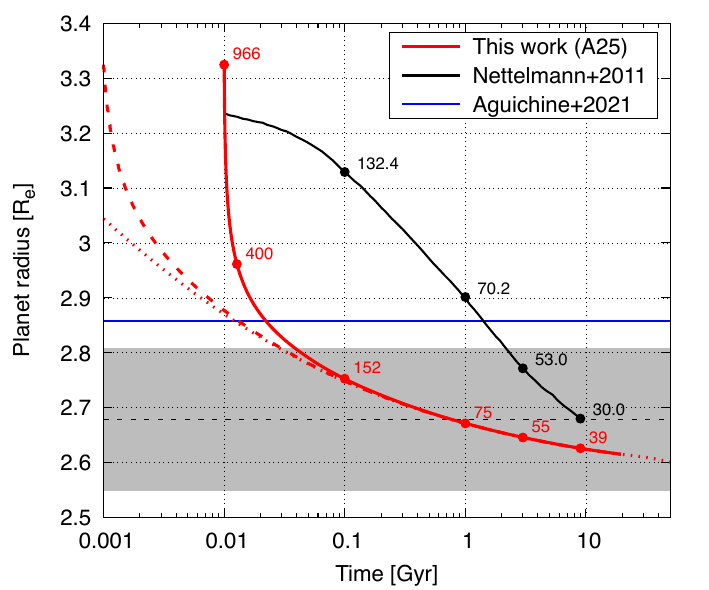}
    \caption{Evolution of the radius of a GJ1214b-like planet of mass 6.55 $\mearth$ with 97\% \ce{H2O} and equilibrium temperature of 555 K orbiting an M-type star. Red, black, and blue lines are this work, case IIb from \cite{Nettelmann2011}, and the static model from \cite{Aguichine2021}. The A21 and A25 models assume $P\sub{tr}=1$ $\mu$bar, and N11 assumes $P\sub{tr}=20$ mbar. The A25 model is shown for $t_0=10$ Myr (solid), 1 Myr (dashed) and 0 Myr (dotted). Points along the curve and labels represent values of $T\sub{int}$ in K. The red dashed line after 20 Gyr shows evolution towards full exhaustion (see Section \ref{sec:coupling-method}). The horizontal dashed black line corresponds to the central value of the radius of GJ 1214b, and the shaded area is the associated uncertainty \citep{Charbonneau2009}.}
    \label{fig:compare-radius}
\end{figure}
\begin{figure}[!h]
    \centering
    \includegraphics[width=0.8\linewidth]{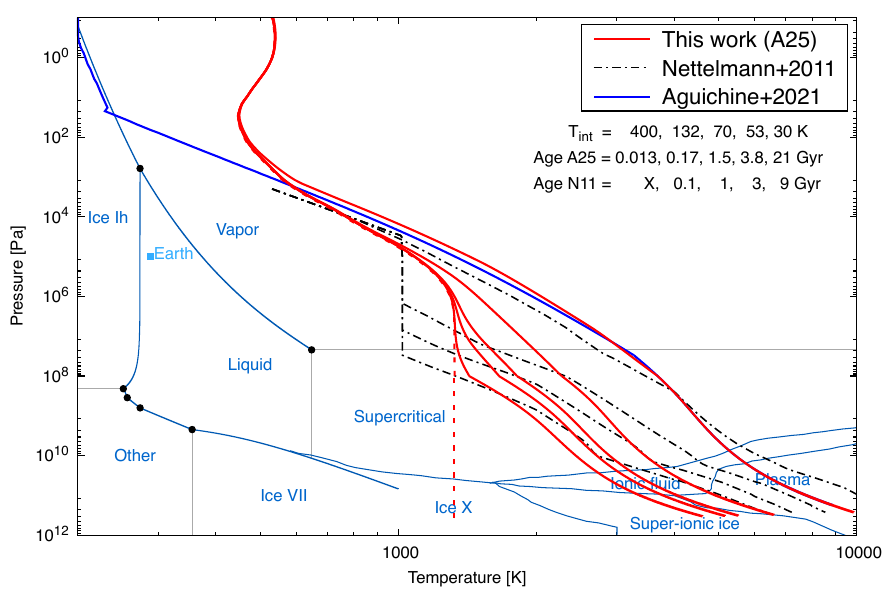}
    \caption{Phase diagram of water with Pressure-Temperature profiles in the planet interior and atmosphere at several given times of the planet evolution, for the planet evolution shown in Figure \ref{fig:compare-radius}. Red lines are the profiles of the A25 model, shown for $T\sub{int}=$ 400, 132, 70, 53, and 30 K. The red dashed line represents the theoretical fully exhausted state of this planet. Black dash-dotted lines are the profiles of the N11 model, shown for $T\sub{int}=$ 132, 70, 53, and 30 K. The corresponding ages for for the A25 and N11 models, for each value of $T\sub{int}$, are given on the figure. The blue line corresponds to the structure computed in \cite{Aguichine2021} for a 5 $\mearth$ planet with $T\sub{eq}=700$ K and $x\sub{H_2O}=1$. This structure does not correspond to the planet radius depicted in Figure \ref{fig:compare-radius}, but is the structure available closest to parameters of GJ 1214b. Phase boundaries are from \cite{Wagner2011} and \cite{Nettelmann2011}.  
    }
    \label{fig:compare-phase}
\end{figure}

Figure \ref{fig:compare-radius} shows the evolution of the planet's radius as a function of time computed by this model A25, with a comparison to the evolutionary model N11 and the static model A21. The A25 model is shown for three initial times $t_0=10$ Myr (solid), 1 Myr (dashed) and 0 Myr (dotted). The shaded rectangle corresponds to the radius of GJ 1214b with its uncertainty \citep{Charbonneau2009}. Figure \ref{fig:compare-phase} shows the corresponding Pressure-Temperature profiles in the planet interior and atmosphere at several given times of the planet evolution, over-plotted with the water phase diagram. As the planet evolves, the interior is cooling, so that the radius decreases over time, and the adiabat of the interior gradually becomes cooler. As the adiabat gets cooler, a secondary isothermal region emerges. It stretches from 10 to 1000 bar at 1326 K for A25, while from 0.1 to 500 bar at 1021 K for N11. We attribute the difference in temperature to the neglection of the collision-induced absorption of H2O in the atmosphere model of \cite{Miller-Ricci2010}, which is used in N11. The interior is fully exhausted when the isothermal region extends to the bottom of the envelope (red dashed line in Figure \ref{fig:compare-phase}), which is not modeled here but is represented for illustrative purposes. The depth to which the secondary isothermal region extends determines the pressure to which the atmosphere profile must be computed, and where the interior model should be connected.


While N11 and A25 cover a similar range of planetary radii: 3.2 to 2.7 $\rearth$ for N11, and 3.3 to 2.6 $\rearth$ for A25; at any given age the radius in A25 is systematically smaller than the radius from N11, and the two evolution curves never cross.
In particular, the A25 evolution curve is marked by a sharp decrease in radius during the first 10 Myr, during which the radius decreases by 15\%, and becomes much slower after 0.1 Gyr, with a change in radius of 5\% between 0.1 Gyr and 20 Gyr. In contrast, N11 appears more uniform as it exhibits a linear trend in log-normal space during most of the evolution, i.e. $R\sub{p}\propto -\log t$.

At any given age, the intrinsic temperature in A25 is higher than the intrinsic temperature in N11, meaning that heat is evacuated faster. In particular, in the A25 model the heat escape rate is $\sim 10^4$ times higher at $t=t_0$, when $T\sub{int}=966$ K, than at $t=0.4$ Gyr, when $T\sub{int}=98$ K, since $L\sub{int}\propto T\sub{int}^4$. This explains the sharp decrease of the radius in A25 at $t=t_0$. We also note that the atmosphere grid of \cite{Kempton2023} only goes up to $T\sub{int}=400$ K, meaning atmospheric properties are extrapolated beyond this value, so that the evolution of the planet radius before 13 Myr is likely approximate. At later stages of the planet evolution, due to the high dependency of $L\sub{int}$ on $ T\sub{int}$, a difference in intrinsic temperature of 25\% will result in a factor of 2 difference in the heat escape rate.

Curves for different initial times give similar radius after $t=0.1$ Gyr, but there is non-negligible difference at early times. The area between both curves correspond to the level of uncertainty in radius due to the choice of initial time $t_0$ when a hot start is assumed. Choosing any arbitrarily high value $s_0$  corresponds to some extreme value of $T\sub{int}$, especially since properties are extrapolated beyond $T\sub{int}=400$ K. This results in the extremely fast evolution of the radius during the first 1 Myr. We therefore choose $t_0 = 1$ Myr so that this extreme behavior does not impact retrievals of planetary bulk composition. With this choice, we conclude that heat loss rates are similar between both models, but in our model the steam envelope is contracting at a slower rate than for the N11 model. This difference most likely comes from the difference in EOS, both in terms of density and thermal gradient, used in both models.

In Figure \ref{fig:compare-phase}, the $(P,T)$ profiles for A25 at $T\sub{int}=400$ K and N11 at $T\sub{int}=132$ K show good visual agreement, and therefore are expect to produce similar radii. However, the radii for these structures are 2.96 $\rearth$ and 3.13 $\rearth$, respectively. When comparing the $(P,T)$ profiles at the same $T\sub{int}=132$ K, 2nd warmest profile of A25 and warmest profile of N11, we find that that the interior part, where $P>1000$ bar, of A25 is significantly colder, and this statement remains true for all values of $T\sub{int}$.
Comparing the density of pure water at similar $(P,T)$ conditions (where the $(P,T)$ profiles of A25 and N11 cross), we find that the density in the N11 model is always $\sim 3\%$ lower than the density in A25. This means that computed planetary radii are intrinsically different by $\sim 1\%$ (since $R\sub{p}\appropto \rho^{1/3} $) because of the choice of EOS alone: the \ce{H2O}-REOS in N11, which combines the SESAME 7150 EOS \citep{Lyon1992} for vapor and supercritical water, and the \cite{French2009} EOS for $T\geq 1000$ K and $\rho \geq 2$ g.cm$^{-3}$, and the \cite{Mazevet2019} EOS in A25.

In the A21 model, atmospheric profiles were computed from the bottom up, starting at a given surface temperature $T\sub{surf}$ and building an adiabat upwards. The RCB was defined as the location where the adiabat and the water condensation curve intersect, where the radiative transfer was computed. The atmosphere and interior were connected at a pressure of 300 bar, where both of them had adiabatic $(P,T)$ profiles. Such a model did not include a secondary isothermal region by construction, which is equivalent, here, to cases of high intrinsic temperature. For this reason, with the same planetary parameters (mass and composition), the static model A21 matches the evolutionary model A25 after $\sim 2.5$ Myr of evolution ($t=12.5$ Myr for the red solid line), when $T\sub{int}\simeq400$ K.

The $(P,T)$ profiles of the A25 model exhibit a $\mathcal{C}^1$ discontinuity at $P\sub{b}=1000$ bar, where the interior and atmosphere models are connected. As mentioned in the beginning of Section \ref{sec:coupling-method}, these kinks are due to the fact that the atmosphere model is not exclusively convective at its bottom, so the thermal gradient is not fully adiabatic. The contribution of collision induced absorption to opacity scales quadratically with pressure, meaning that any non-exhausted interior is expected to have a finite secondary isothermal region, and become fully adiabatic at some pressure $P\sub{ad}$, given that the envelope is thick enough to reach that pressure. \cite{Kempton2023} re-calculated the opacity of \ce{H2O} up to 1000 bar so that as many atmosphere profiles as possible are adiabatic at the atmosphere's base. However, our coupled modeling shows that this is satisfied only when $T\sub{int}>100$ K, in which case we underestimate $P\sub{ad}$. Even in the approximation of radiative diffusion, computing the thermal gradient would require the mean Rosseland opacity at pressures greater than 1000 bar, which is currently not available and would require more experimental data and dedicated studies. Extrapolating the mean Rosseland opacity is not desirable due to its complex and non-linear behavior at high pressure and temperature. The consequence of this limitation is that the entropy of the envelope (i.e. the temperature of the interior) is overestimated, meaning the radius is slightly overestimated. The error on the radius can be roughly estimated by computing the structure of the fully exhausted interior -- when the interior of the planet is entirely isothermal (dashed line in Figure \ref{fig:compare-phase}). For our test case, shown in Figure \ref{fig:compare-radius}, at 20 Gyr the radius is 2.61 $\rearth$, and $T\sub{int}=31$ K. At this rate, it would take $\sim 230$ Gyr to achieve full exhaustion, and the planet reaches a radius of 2.52 $\rearth$. This corresponds to a barely measurable change in radius over a timescale much larger than the typical age of exoplanets. We therefore conclude that $P\sub{b}=1000$ bar is sufficient for general demographic studies of steam-worlds around M-stars, but an update will be required for dedicated studies when the uncertainty on planetary radius is less than $5\%$.

Another difference arising from the choice of the EOS is the difference in the thermal gradient, determined by the adiabatic exponent $\nabla\sub{ad}$ (or, equivalently, the Grüneisen parameter $\gamma$). The thermal gradient is always greater in the N11 model than the A25 model: even when adiabats cross (see, e.g. $P=1000$ bar and $T=2100$ K in Figure \ref{fig:compare-phase}), the temperature is rising faster in the N11 model than A25. In the N11 model, water at the bottom of the envelope is always in plasma state, whereas in the A25 model, super-ionic ice begins to form at the bottom envelope when $T\sub{int}=132$ K. Since the temperature in the interior is higher for the N11 model, the density will be lower, so that the computed planet radius will be greater. We conclude that warmer adiabats in the deep interior of the N11 model contribute even more to produce a greater planetary radius.

In summary, we identified 3 key difference between the N11 and the A25 model:
\begin{itemize}
    \item Both models have similar heat loss fluxes, but the envelope is contracting at a slower rate in the A25 model, which uses the \cite{Mazevet2019} EOS, than in the N11 model, which uses the \ce{H2O}-REOS \citep{Nettelmann2008}.
    \item The density of water is $\sim 3\%$ higher when computed with the \cite{Mazevet2019} EOS (model A25) than with the \ce{H2O}-REOS \citep{Lyon1992,French2009} (model N11), resulting in $\sim 1\%$ smaller planets.
    \item The thermal gradient is greater when computed with the \ce{H2O}-REOS \citep{Nettelmann2008} than \cite{Mazevet2019} EOS, meaning the interior is warmer in the N11 model than in the A25 model. This effect leads to a 10\% larger radius in the N11 model for GJ 1214b over much of the evolution. 
\end{itemize}
The 2 latter effects combine in the same direction, resulting in smaller planetary radii for the A25 model, compared to the N11 model. We also identified that, by construction, the A21 model will often match the radius of A25 at early ages. We note that in the domain of high pressures and temperatures, both the \ce{H2O}-REOS and \cite{Mazevet2019} EOS were constructed from ab initio simulations. This highlights the difficulty of choosing between different EOS, since a careful a thorough analysis must be performed to assess the accuracy of the ab initio simulations.

These differences are significant enough to impact our conclusions when interpreting the bulk composition of sub-Neptunes. Our analysis shows that A21 tends to underestimate the WMF, since the computed radius is overestimated, as most planets are old. The A25 model is consistent with the radius of GJ 1214b at $1-\sigma$ between 7 Myr and 160 Gyr. This means that if sub-Neptunes are pure steam worlds, and their radius is known with a precision better than $\sim 5\%$, measuring the age of an individual planet would not provide an additional constraint on its composition. At a population level, a measured absence of a correlation between radius and age in a sample of sub-Neptunes could indicate that they are pure steam worlds \citep[as illustrated in Fig. 16 of][]{Rogers2021}. Conversely, the presence of a correlation between radius and age would indicate that they are not pure steam worlds but a mixture of volatiles and/or \ce{H2}-He.


\subsection{New insights on sub-Neptunes' interiors}

\begin{figure}[!h]
    \centering
    \includegraphics[width=0.49\linewidth]{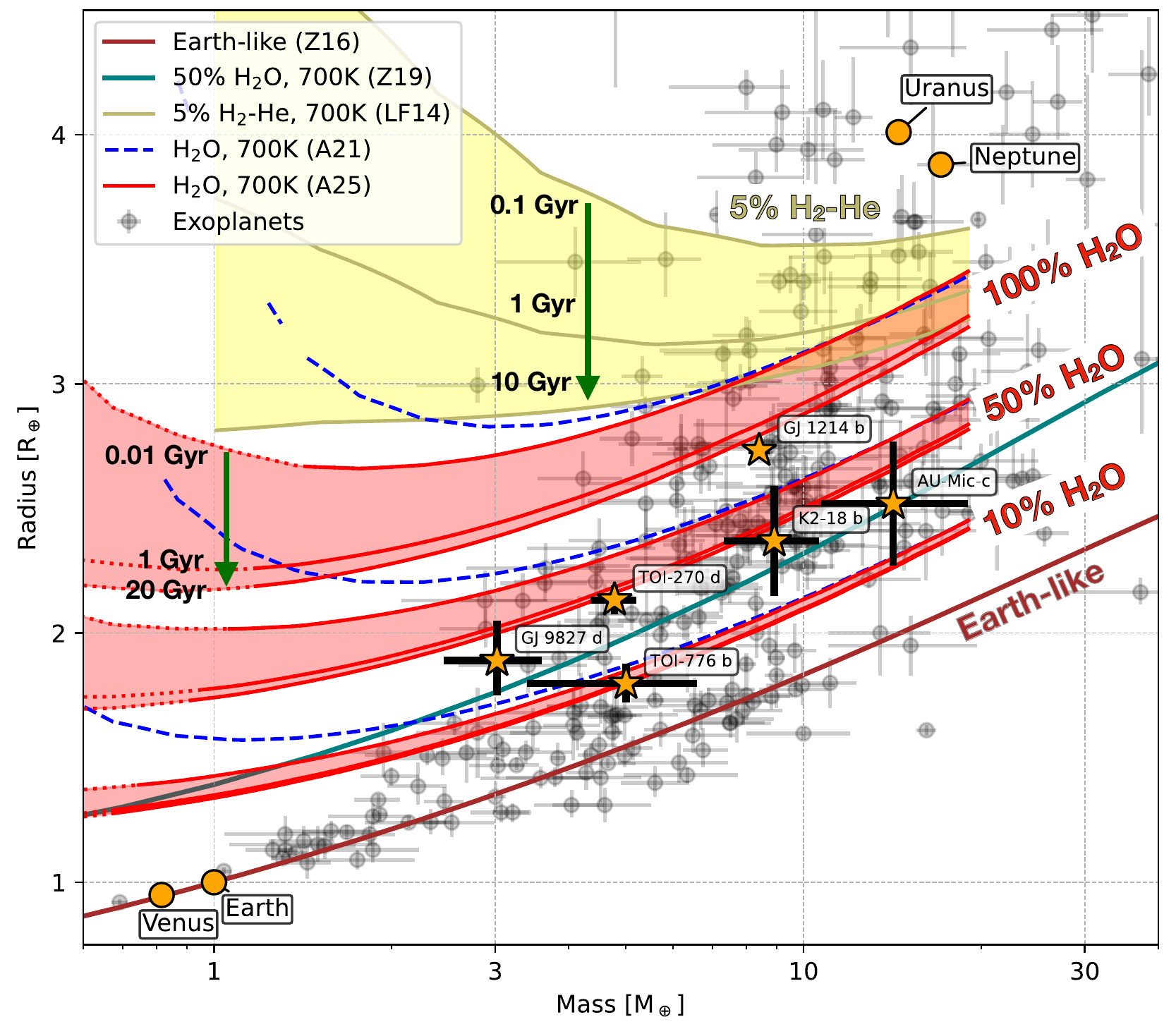} \includegraphics[width=0.49\linewidth]{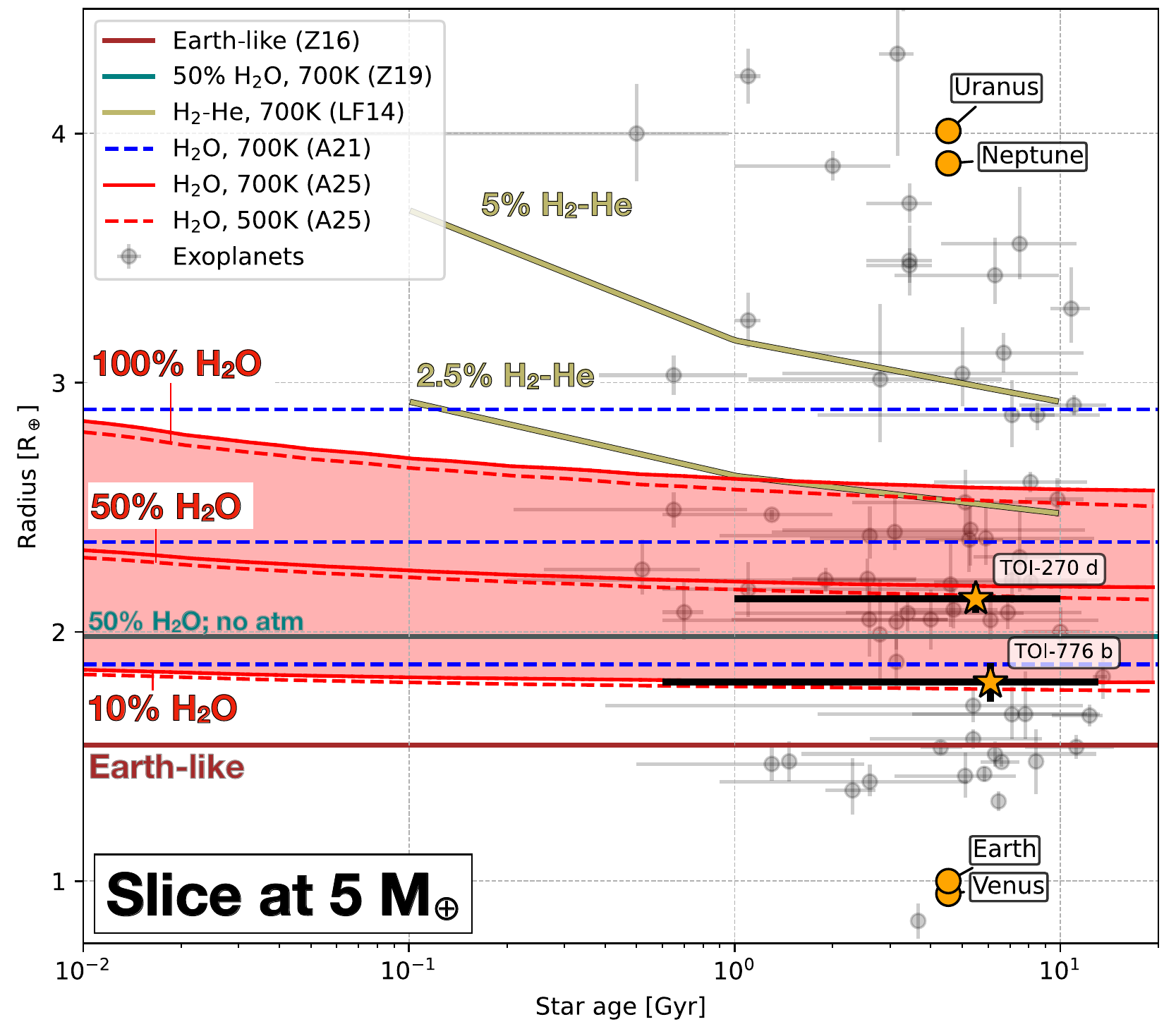}
    \caption{Interior structure models in the mass-radius plane (left panel) and slice at $5~\mearth$ in the radius-time plane (right panel): evolving steam worlds (red solid lines, A25), static steam worlds (blue dashed lines, A21), 50\% isothermal steam worlds (teal solid line, Z19), Earth-like planets (brown solid line, Z16), and 5\% \ce{H2}-He planets (yellow solid lines, LF14). Red and blue lines are shown for bulk water content of 10\%, 50\% and 100\%, and for the red lines an M-type host star is assumed. Red and yellow shaded regions represent the span of evolution models of A25 and LF14, respectively. LF14 evolution is shown between 0.1 and 10 Gyr. A25 evolution is shown between 0.02 Gyr and 20 Gyr. All models assume an equilibrium temperature of 700 K, except the red dashed line in the right panel, which shows the A25 model at 500 K. Red dotted lines correspond to cases where the steam atmosphere thickness is underestimated by $1\%$ because of constant surface gravity (see Appendix \ref{sec:appendix-constant-g}). Exoplanets properties are obtained from the NASA Exoplanet Archive, including 6 planets of interest shown with orange stars (see Appendix \ref{sec:appendix-catalog} for details). Solar system planets are shown with orange circles.}
    \label{fig:mass-radius-age}
\end{figure}

In Figure \ref{fig:mass-radius-age}, left panel, we shows the mass-radius relations for the A25 model, the A21 model, and the 50\% \ce{H2O} line from \cite{Zeng2019} noted Z19. \cite{Zeng2019} provide mass-radius curves for 50\% and 100\% \ce{H2O} planets for temperatures between 300 K and 1000 K. The Z19 model assumes an isothermal \ce{H2O} layer with a surface pressure of 1 mbar. The \cite{Zeng2016} (Z16) model provides a grid of planet radii as a function of mass and composition, but only at 300 bar of surface pressure. We use the Z16 model to infer WMF. For planets older than 20 Myr, the planetary radius computed with the A25 model is smaller than in the A21. However, the Z19 model predicts even lower radii, due to the absence of thermal gradient. The red dotted lines represent the parameter space where the error on the atmosphere's thickness due to non-constant surface gravity $\epsilon\sub{grav}$ is greater than 1\% (see Eq. \ref{eq:epsilon-gravity} in Appendix \ref{sec:appendix-constant-g}). Since this only occurs in a parameter space where no exoplanets have been detected yet, we conclude that using a constant surface gravity in the atmosphere model is sufficient for steam atmospheres.
The right panel of Figure \ref{fig:mass-radius-age} corresponds to a slice of the mass-radius plane at $5~\mearth$, showing the radius as a function of planet age. Planets with \ce{H2}-He envelopes exhibit a moderate change in radius with age, as found by \cite{Lopez2014}, noted LF14. Our results show that radii of steam worlds have a very weak dependence on age, except at very young age, during the first 1 Myr of their evolution. This implies that planet age will not help further constrain the bulk water content of an individual steam world unless the planet is extremely young (few Myr). However, this also means that age is not critical to infer the bulk water content of steam worlds. Interestingly, we notice that the 50\% \ce{H2O} line of the A25 model coincides with the model of the sub-Neptune distribution above the radius valley between 2 and 8 Gyr at $\sim2.2~\rearth$. This may suggest that this population of exoplanets are volatile-rich. However, the sample is too small to perform a quantitative analysis.

While Figure \ref{fig:compare-radius} shows results of our model only assuming a M-type host star, our grid also computes the evolution of steam worlds around a Sun-like host star. Since \ce{H2O} is a strong greenhouse gas, energy from a Sun-like star will be deposited deeper in the planet than energy from a M-type host. The first effect is that the interior of planets around Sun-like stars will be warmer, and their radius will be greater. A second effect is that due to the transparency of \ce{H2O} to high energy radiation, planets around Sun-like stars are closer to radiative equilibrium, which translates to smaller values of $T\sub{int}$ than for planets around M-type hosts. As a consequence, the radius of planets around Sun-like stars contracts slower. These effects also depend on the column density of water that energy is radiating through, so we restrict this analysis to $M\sub{p}\ge 1$ $\mearth$. At early ages ($\sim 20$ Myr), the radius is mostly determined by the initial entropy of the envelope, and the difference in radius between both cases is less than $0.5\%$. At 20 Gyr, the radius of planets around Sun-like stars is up to 8\% greater than planets orbiting M-type stars, when $M\sub{p}\ge 1$ $\mearth$.\\

\begin{figure}[!h]
    \centering
    \includegraphics[width=0.50\linewidth]{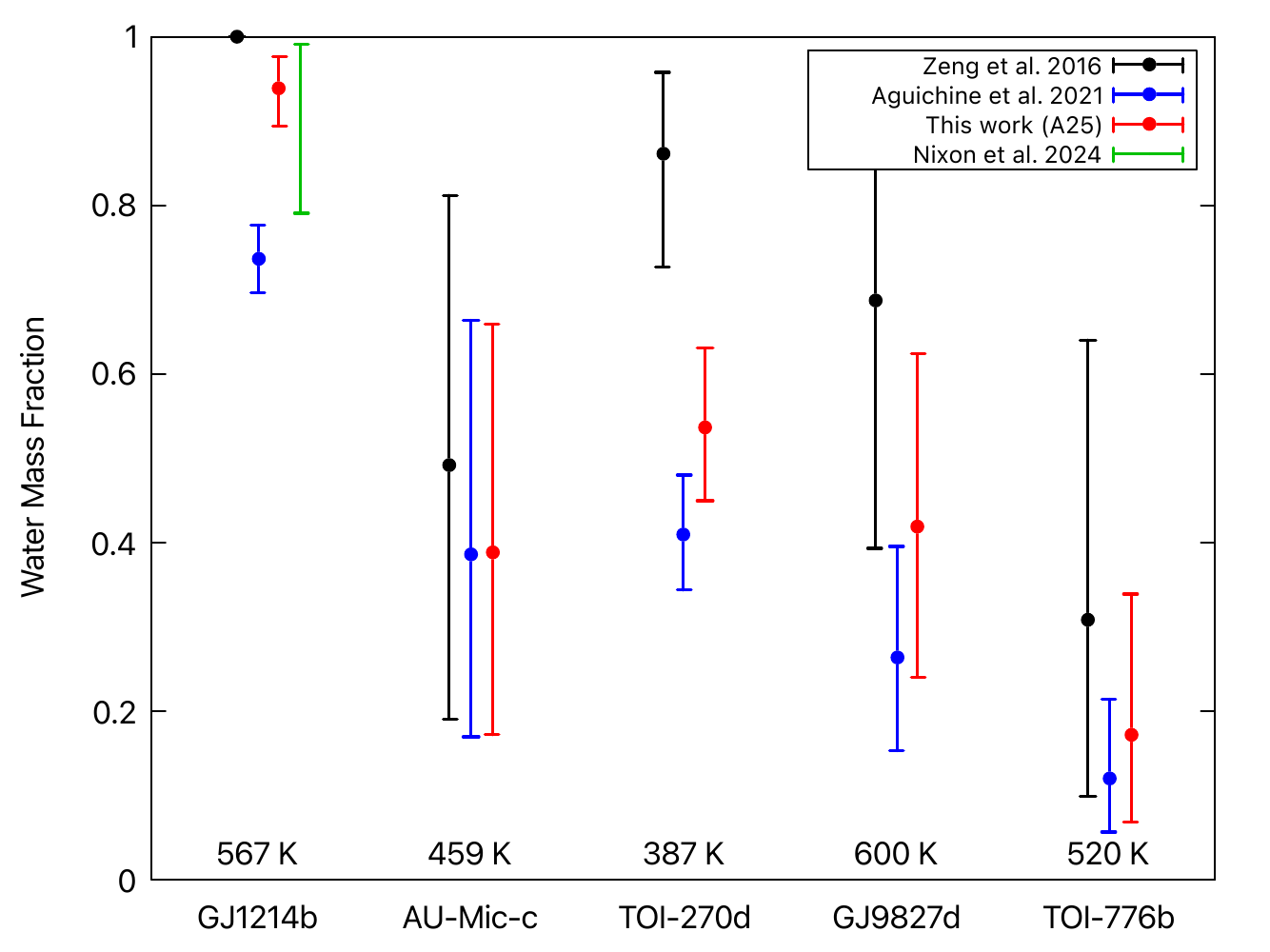}
    \caption{Inferred water mass fraction of 5 exoplanets, 4 of which have been observed with JWST and found to have high mean molecular weight atmospheres. The WMF were inferred using the Z16 \citep{Zeng2016}, A21 \citep{Aguichine2021} and A25 (this work) models. Green values correspond to the range of WMF inferred by the model of \cite{Nixon2024}. Planets were ordered by decreasing central value of planetary radius.}
    \label{fig:wmf}
\end{figure}

The most important characteristic of an interior model is the reinterpretation of the bulk composition of exoplanets. We compare the bulk water content inferred from different interior models from the literature for 5 planets: TOI-776 b, GJ 9827 d, TOI-270 d, AU Mic c and GJ 1214 b. For GJ 1214 b, we now use the latest update on the planet properties from \cite{Mahajan2024}. All planets except AU Mic c have been observed with JWST, and have been found to have high mean-molecular weight atmospheres \citep{Kempton2023-gj1214b,Piaulet-Ghorayeb2024,Benneke2024,Teske2025}. AU Mic c is the youngest sub-Neptune of the exoplanet catalog, with a stellar age of $20.1^{+2.5}_{-2.4}$ Myr. 
K2-18b has been excluded from this analysis because of its low $T\sub{eq}=255$ K, and observations in transit show that it has a \ce{H2}-He dominated escaping atmosphere \citep{dosSantos2020,Madhusudhan2023}. To infer the bulk water content of each planet, we adopt a Bayesian approach using the \texttt{emcee} sampler \citep{Foreman-Mackey2013}. For each planet, we adopt gaussian priors on mass and equilibrium temperature, and a gaussian or flat prior on age, depending on available measurements. The likelihood function is a gaussian that compares the sampled radius, computed from the model, to the measured radius. The water mass fraction (WMF) is assigned a flat prior from 0 to 1, and the posterior corresponds to the distribution of possible WMF in the planet. We compare three models: A25 (this work), A21 (previous work), and Z16 \citep[isothermal \ce{H2O} envelope][]{Zeng2016}. The origin of planetary parameters is given in Appendix \ref{sec:appendix-catalog}, and our results are summarized in Figure \ref{fig:wmf}. As expected, Z16 consistently gives the highest estimate of the WMF, and A21 the lowest. These results quantify the difference in inferred WMF when different assumptions are made for the thermodynamic state of the interior. We argue that out of the three models, A25 corresponds to the most realistic thermodynamic state in the interior.

In Figure \ref{fig:wmf} we also added the WMF inferred by the model developed by \cite{Nixon2024}. Their coupled atmosphere and interior model, however, does not evolve in time. Instead, they assume that the planet is in a thermodynamic state corresponding to $T\sub{int}=30$ K. Furthermore, they connect the atmosphere model to the interior at 100 bar. Since the depth of the secondary isothermal region determines the entropy of the deep adiabat (see Fig. \ref{fig:compare-phase}), the interior will be warmer if connected at 100 bar compared to 1000 bars in our model. This explains why \cite{Nixon2024} find lower possible WMF for GJ 1214b. As seen in Figure \ref{fig:compare-phase}, for values of $T\sub{int}$ lower than 100 K, it seems that even 1000 bar is not sufficient to capture the extent of the secondary isothermal region, meaning that our model also slightly overestimates the radius. This highlights the difficulty of modeling realistic interiors of sub-Neptunes and the limitations caused by the lack of opacity tables at such high pressure and low temperature conditions. Part of the uncertainty in the value inferred by \cite{Nixon2024} comes from considering a range of haze production rates, changing the transiting radius of the planet. When the haze production rate is low, the atmosphere is clear, so that the transiting radius is at a lower altitude, and results in a higher inferred WMF.

For the three models used in Fig.~\ref{fig:wmf}, despite central values being clearly distinct, there is a considerable overlap in inferred WMF uncertainty. This is especially true for AU Mic c, which has a huge uncertainty in both mass and radius. AU Mic c is also the model where A21 and A25 yield similar results, further confirming that the A21 model behaves as an early-age version of A25. Significant overlap between models is also observed for TOI-776, due to its large uncertainty in mass, though to a lesser extent thanks to the better precision on radius. In contrast, the WMF of TOI-270d is reasonably well constrained, which has its mass measured to high precision (9\% error) with the Echelle Spectrograph for Rocky Exoplanet and Stable Spectroscopic Observations (ESPRESSO) and the High Accuracy Radial velocity Planet Searcher (HARPS). This result highlights the necessity for better mass and radius measurements to properly characterize bulk compositions, especially in an era of exoplanet science with a wide diversity of interior models.\\

All evolution tracks produced in this work are available on Zenodo \dataset[doi:10.5281/zenodo.15043384]{https://doi.org/10.5281/zenodo.15043384} \citep{Aguichine2025-swe-zenodo} for graphical representation purposes, and can be used to quantitatively infer the bulk water content in sub-Neptunes, e.g. using MCMC. Evolution tracks also provide quantities of interest such as the specific entropy of the envelope, contributions to the total heat loss rate from the core, envelope and radiogenic heating in the form of luminosities, and the moment of inertia factor $I/(M\sub{p}R\sub{p}^2)$. For visualization in the mass-radius plane, the evolution tracks have been implemented in \texttt{mardigras} v2.3.1\footnote{\href{https://github.com/an0wen/MARDIGRAS}{https://github.com/an0wen/MARDIGRAS}} \citep{Aguichine2024-mardigras,Aguichine2025-mardigras-v2}.

\section{Conclusion} \label{sec:ccls}

In this work, we computed the time evolution of Steam Worlds using an updated coupled atmosphere and interior model to better understand how composition affects the observed properties of sub-Neptunes. This framework combines an interior model of an Earth-like core with a pure water envelope \citep{Aguichine2021}, and a pure steam atmosphere on-top \citep{Kempton2023} self-consistently. The evolution is computed by integrating the energy equation, assuming the accretion energy during formation results in a so-called "hot start" \citep{Fortney2004PhD,Lopez2012}. We compared our work to a model with similar physics presented in \cite{Nettelmann2011}, noted N11, and found that our model predicts smaller planetary radii than the N11 model, as shown in Figure \ref{fig:compare-radius}. We identified three main differences: i) the envelope is contracting at a slower rate in the A25 model, which uses the \cite{Mazevet2019} EOS, than in the N11 model, which uses the \ce{H2O}-REOS \citep{Nettelmann2008}, ii) the EOS used here \citep{Mazevet2019} predicts a $\sim 3\%$ higher density than the EOS in the N11 model given the same pressure and temperature conditions, and iii) the thermal gradient in this model is less strong, further increasing the density when compared to the N11 model. Despite using different atmospheric compositions for radiative heat transport (pure steam for A25, $50\times$ Solar for N11), both models have similar heat loss fluxes from the interior. This model was also compared with previous work \citep[][noted A21]{Aguichine2021}, showing that the physics used to construct the A21 model is representative of younger planets (see Figure \ref{fig:compare-phase}) and therefore overestimates the radius of older planets. On the other hand, interior models with isothermal \ce{H2O} envelopes such as \cite{Zeng2016,Zeng2019} underestimate the radius significantly. A grid of all evolutionary tracks are available on Zenodo \dataset[doi:10.5281/zenodo.15043384]{https://doi.org/10.5281/zenodo.15043384} \citep{Aguichine2025-swe-zenodo}, although we warn users about the limited validity range in equilibrium temperature space.\\


We have identified similar behavior of \ce{H2O} envelopes as is known for \ce{H2}-He envelopes regarding the hot start or cold start choice \citep{Marley2007,Spiegel2014}. A hot start can result in extremely fast, even nonphysical, decrease in radius during the first few Myr of the planet evolution. After this initial stage, the change in radius slows down and becomes smaller than the typical uncertainty on measured planetary radii. This means that age does not help further constraining the bulk water content of steam worlds, even for the youngest known sub-Neptune AU Mic c \cite[$20.1^{+2.5}_{-2.4}$ Myr][]{Wittrock2023}. From an observational perspective, if sub-Neptunes exhibit a correlation between radius and age, they are likely not pure steam worlds but their envelope is a mixture of volatiles (most likely \ce{H2}-He). Conversely, the absence of correlation between sub-Neptunes radii and ages may indicate the existence of pure steam worlds.\\

One interesting observation is the possible transition of \ce{H2O} in the deep interior from the plasma state to the super-ionic ice state (Fig. \ref{fig:compare-phase}). In super-ionic \ce{H2O} ice, oxygen atoms form a solid lattice and only hydrogen atoms can diffuse in a liquid-like fashion \citep{Demontis1988}. The resulting change in viscosity and electrical conductivity \citep{Millot2018,Millot2019} is likely to affect the dynamo in the planet \citep{Tian2013} and the thermal evolution through reduced vigor of convection \citep{Stixrude21}. For the case of GJ 1214b, this phase transition begins at the bottom of the hydrosphere at $0.42$ Gyr (see the red line $T\sub{int}=100$ K in Figure \ref{fig:compare-phase}). 

Our new set of mass-radius relationships (left panel of Fig. \ref{fig:mass-radius-age}) implies a re-interpretation of the bulk composition of sub-Neptunes. For older planets, the previous model \citep{Aguichine2021} consistently overestimates the radius, and hence underestimates the bulk water content. On the other hand, the canonically used \cite{Zeng2016} model consistently underestimates the radius, and hence overestimates the bulk WMF. We also note that for most planets, there is still significant overlap in inferred WMF due to the large uncertainty on planet mass and radius. If the bulk WMF is known (from atmospheric measurements or theoretical predictions), in order to discriminate between interior structure models one needs mass and radius uncertainty comparable to that of TOI-270 d, respectively 19\% and 8\%. To reach this precision level, we advocate for more radial velocity observations using high precision facilities to constrain mass, e.g. with HARPS and ESPRESSO, and more transits using high sensitivity instruments to constrain radius, such as JWST, PLATO, and CHEOPS.\\

We emphasize that in this model, the envelope and atmosphere are made of pure \ce{H2O}, which is an idealized end-member case of a planet that formed without accreting any \ce{H2}-He gas, and where \ce{H2O} is a proxy for all volatiles. The main effect expected from the addition of \ce{H2}-He is a decrease in density that will significantly increase the radius of the planet. Despite its high opacity, \ce{H2O} has absorption bands and absorption gaps. The addition of volatiles such as \ce{CH4} and \ce{NH3} has the potential to fill these gaps, increasing the mean Rosseland opacity of the mixture. We speculate that such an increase in mean Rosseland opacity could slow down the cooling of the interior, and thus the contraction of a planet. However, changing the composition of the atmosphere will also require a change in the EOS in the interior, which would result in a completely new class of models. Effectively, Section \ref{sec:results-compare} compares a model of pure steam envelope with a pure steam atmosphere (A25) to a model of pure steam envelope with a $50\times$ Solar atmosphere. We find minimal differences in terms of heat escape rate, as indicated by values of $T\sub{int}$, despite the fact that \ce{H2}-He collision induced absorption is expected to dominate at high pressures, and fill absorption gaps. Although, the addition of \ce{H2}-He to our model constitutes the next step in the development of this coupled model. Additional volatiles  (\ce{CH4}, CO, \ce{CO2}, \ce{NH3}, etc.) will be the subject of future work. Our model also does not consider mixture of \ce{H2O} with the core. This mixing can be partial \citep{Dorn2021, Luo2024} or total \citep{Vazan2022}. Mixture of \ce{H2O} with the material in the core results in a smaller planet radius, further increasing the inferred WMF.\\

Although a pure steam atmosphere and envelope is a highly idealized case, this study showcases the main differences between the time evolution of \ce{H2}-dominated planets \citep{Lopez2014,Tang2024} and steam worlds (this work), which is necessary to understand sub-Neptunes on a population level. We believe that this model pushes forward our understanding of planetary atmospheres and interiors, and refines the inference of the composition of sub-Neptunes. This study unveils possible links between the bulk composition of sub-Neptunes and the system's age, both of which are being actively measured by new missions such as JWST and Gaia, and in the future HWO, ARIEL, and PLATO.
\begin{acknowledgments}
This material is based upon work supported by NASA’S Interdisciplinary Consortia for Astrobiology Research (NNH19ZDA001N-ICAR) under award number 80NSSC21K0597. JJF acknowledges the support of NASA Exoplanets Research Program grant 80NSSC24K0691. JEO is supported by a Royal Society University Research Fellowship. This project has received funding from the European Research Council (ERC) under the European Union’s Horizon 2020 research and innovation programme (Grant agreement No. 853022).  This research has made use of the NASA Exoplanet Archive, which is operated by the California Institute of Technology, under contract with the National Aeronautics and Space Administration under the Exoplanet Exploration Program. This work has benefited from the 2023 Exoplanet Summer Program in the Other Worlds Laboratory (OWL) at the University of California, Santa Cruz, a program funded by the Heising–Simons Foundation.
\end{acknowledgments}

%



\software{\texttt{NumPy} \citep{Harris2020-numpy}, \texttt{matplotlib} \citep{Hunter2007-matplotlib}, \texttt{scipy} \citep{Virtanen2020-scipy}, \texttt{emcee} \citep{Foreman-Mackey2013}, \texttt{astropy} \citep{astropy:2013,astropy:2018,astropy:2022}, \texttt{mod\_fint} \citep{mod_fint2024}.}



\appendix

\section{Numerical scheme for time evolution} \label{sec:appendix-evolution-scheme}

The numerical procedure for our integration is similar to the one presented in \cite{Fortney2004PhD}. From any given state where the entropy of the envelope is $s$, a new state is generated with an entropy $s-\Delta s$. Both states will have a difference in interior energy of $\Delta E_{\mathrm{int}}$. Therefore, the time interval $\Delta t = \Delta E_{\mathrm{int}} / <L_{\mathrm{out}>}$ that separates both states can be computed, where $<L_{\mathrm{out}}>$ is the average luminosity escaping from the planet during the time step. Since some energy is produced in the interior by radioisotopic decay, $L\sub{out} = L\sub{int} - L\sub{radio}$.

For a given value of specific entropy, the surface temperature $T_{\mathrm{b}}$ is obtained by inverting the equation of state of water, knowing $P_{\mathrm{b}}=1000$ bar. The interior model is then run with the provided composition ($x_{\mathrm{core}}$, $x_{\mathrm{H_2O}}$), mass $M_{\mathrm{b}}$, and surface conditions ($P_{\mathrm{b}}$, $T_{\mathrm{b}}$). This provides the planet radius and surface gravity. The atmosphere grid $T_{\mathrm{b}}(\log g,T_{\mathrm{int}})$ can then be inverted to obtain the $T_{\mathrm{int}}$ that corresponds to this structure, which is used to compute $L_{\mathrm{int}}$.

\section{Atmospheric thickness and surface gravity} \label{sec:appendix-constant-g}

In this section, our goal is to derive a quantitative criterion on the validity of the constant gravity assumption. For an atmosphere that is isothermal at a temperature $T_0$ and in hydrostatic equilibrium with a surface gravity $g_0$, the atmospheric scale-height is $H = k\sub{B}T_0/(\mu g)$, where $\mu$ is the mean molecular weight of the atmosphere, and the ideal gas law is assumed to apply. If the gravity is assumed constant with altitude $z$, the pressure profile is \citep[see e.g. appendix of][]{Turbet2020}:
\begin{equation}
    P(z) = P_0 \exp \left( -\frac{z}{H}\right),
\end{equation}
where $P_0$ is the pressure at the surface level. The thickness from $P_0$ to the transit pressure $P\sub{tr}$ is then $Z_1 = \alpha H$, where we note $\alpha = \ln \left( P_0/ P\sub{tr}\right)$. If gravity is varying with height, the pressure profile is instead:
\begin{equation}
    P(z) = P_0 \exp \left( \frac{R\sub{0}}{H} \left( 1 - \frac{R\sub{0}}{R\sub{0}+z} \right) \right),
\end{equation}
where $R_0$ is the planet radius at the surface. In this case, the atmospheric thickness is:
\begin{equation}
    Z_2 = \frac{\alpha H}{\alpha H/R_0 -1},
\end{equation}
using the same notations as above. In this case, the relative error on the total planetary radius $R_0 + Z\sub{1 ~or~ 2}$ when using $Z_1$ or $Z_2$ can be estimated as:
\begin{equation}
    \epsilon\sub{grav} = \frac{(R_0+Z_2)-(R_0+Z_1)}{\frac{1}{2}\left[(R_0+Z_2)+(R_0+Z_1)\right]} \lesssim \frac{Z_2-Z_1}{R_0+Z_1} = \frac{\left(\alpha H/R_0\right)^2}{1-\left(\alpha H/R_0\right)^2}. \label{eq:epsilon-gravity}
\end{equation}
In the thin atmosphere approximation $H \ll R_0$, so that $\epsilon\sub{grav} \simeq \left(\alpha H/R_0\right)^2$ which is $\ll 1$ by definition. By inverting Eq. (\ref{eq:epsilon-gravity}), we find that a 5\% difference in radius is achieved when $H/R_0 \simeq 0.02$ (using $\alpha \simeq 10$, see below). We note that since the Jeans parameter for an atmosphere is $\Lambda = R_0/H$ \citep{Jeans1926,Owen2016,Cubillos2017,Vivien2022}, planets where  $\epsilon\sub{grav}\ge 5\%$ also have $\Lambda \le 50$, meaning that most hydrogen dominated atmosphere need to include a non-constant gravity treatment. Since $\Lambda$ is inversely proportional to the mean molecular weight of the atmosphere, steam atmosphere, or other heavy volatile atmospheres, can remain in the thin atmosphere approximation (constant surface gravity). For GJ-1214b, using parameters computed by \cite{Mahajan2024}, we find $\epsilon\sub{grav} \simeq 2\times 10^{-4}$, assuming that the thin atmosphere is between $P_0\sim10^3$ Pa and $P\sub{tr}=0.1$ Pa, and that the molar mass of the atmosphere is 18 g.mol$^{-1}$ (pure water).


\section{Curated exoplanet catalog} \label{sec:appendix-catalog}

Since the composition of exoplanets is interpreted from their mass, radius, temperature and age, it is essential to only represent planets for which these properties were measured properly. The catalog of exoplanets shown in Figure \ref{fig:mass-radius-age} was exported from the NASA Exoplanet Archive using the following filters:
\begin{enumerate}
    \item Default Parameter Set to 1.
    \item Controversial Flag Set to 0.
    \item Both masses and radii are measured and have uncertainty.
    \item Actual Mass Representation (i.e. not Msini or Msini/sini) measured by Radial Velocities.
    \item Mass Uncertainty smaller than 50\%, since large error bars do not help interpretation.
\end{enumerate}
We note that item 4 also filters out all planets from the TRAPPIST-1 system, the masses of which were obtained transiting time variations (TTV).\\

We then select a six exoplanets of interest to characterize their bulk interiors, ordered by increasing planetary radius:
\begin{itemize}
    \item TOI-776b \citep{Fridlund2024}, a radius-gap planet with an inferred atmospheric metallicity of at least 180 times the solar value (atmospheric mean molecular weight of at least 6 g/mol) using JWST transit spectroscopy (Teske et al. 2024, in prep).
    \item GJ 9827d \citep{Piaulet-Ghorayeb2024}, a mini-Neptune found to have a \ce{H2O} dominated atmosphere with a volume mixing ratio of at least 31\% using JWST transit spectroscopy, confirming its status as a Steam World.
    \item TOI-270d \citep{Van-Eylen2021}, a mini-Neptune with signatures of \ce{CH4}, \ce{CO2} and \ce{H2O} detected in its atmosphere by JWST using transit spectroscopy, also having a high mean molecular weight \citep[$\sim 5.5$ g/mol][]{Benneke2024}. For this planet, we use a flat prior on age between 1 and 10 Gyr \citep{Benneke2024}.
    \item K2-18b \citep{Sarkis2018}, a mini-Neptune with signatures of \ce{CH4} and \ce{CO2} detected in its atmosphere by JWST using transit spectroscopy \citep{Madhusudhan2023}, and hypothecized to be a so-called "Hycean World" \citep{Madhusudhan2020}. For this planet, we use the age estimated by gyrochronology from \citep{Guinan2019}.
    \item AU Mic c, the youngest confirmed sub-Neptune with a stellar age of $20.1^{+2.5}_{-2.4}$ Myr. The stellar age and radius are from \cite{Wittrock2023}, and the mass was measured by RV in \cite{Donati2023}.
    \item GJ 1214b \citep{Mahajan2024}, a mini-Neptune with a JWST phase curve that indicates a very reflective atmosphere (clouds or haze) and a high metallicity, despite the lack of molecular features \citep{Kempton2023-gj1214b}.
\end{itemize}



\bibliography{bibliography}{}
\bibliographystyle{aasjournal}



\end{document}